# Establishing and assessing adaptive parking prices in a city: Algorithms, software and examples

Nir Fulman, Itzhak Benenson

Department of Geography and Human Environment, Porter School of Environmental Science, Tel Aviv University, Israel

**Abstract**:

We propose ParkSage, a set of spatially-explicit algorithms for establishing parking prices that guarantee a predetermined occupancy rate over a city, and for evaluating the achieved reduction in parking search time. We apply ParkSage for establishing overnight parking prices that guarantee 85% occupation in the Israeli city of Bat Yam. Pricing by street links ensures high parking availability and close to zero cruising everywhere in the city, but is inconvenient for drivers. Establishing prices by the large and heterogeneous city quarters results in local mismatch between demand and supply, the emergence of areas with fully occupied on-street parking and a long search time for the drivers whose destinations are in these areas. We demonstrate that pricing by the medium sized Transportation Analysis Zones, which is easy enough for drivers to comprehend and abide by, is sufficient for eliminating cruising. The software for establishing and assessing performance parking prices is based on the standard municipal GIS layers of streets and parking lots and is available for free download from https://www.researchgame.net/profile/Nir_Fulman

## 1 Introduction

Every city in the world has its own parking policy. In some cities, on-street parking is provided for free, whereas in most cities in the western world parking costs money. Typically, a large area centered around the CBD is priced uniformly throughout the day. However, demand for parking is determined by the attractiveness of single destination buildings and thus varies substantially in space and time. Underpriced parking in neighborhoods where demand exceeds supply entails long searches for parking, which increase traffic congestion. The associated external costs in terms of time, fuel and environmental degradation can be very high. Overpriced parking, on the other hand, deters drivers from visiting an area and may harm local economic activities (Arnott & Inci, 2006; Pierce & Shoup, 2013).

Studies suggest that keeping about 1 in 7 parking places vacant in each block, or 85% occupancy, would allow drivers to find parking quickly and at the same time be sufficient for maintaining economic activities in the area (Shoup, 2006, Levy et al., 2013). This view has become popular and in recent years, a number of cities around the world have initiated programs for enforcing a predefined level of parking occupation using prices as a control parameter (LADOT, 2019; SDOT, 2019; SFMTA, 2014; AT, 2019; CPA, 2017; City of Boston, 2018; DDOT, 2019; Parking Authority of Baltimore City, 2019; Berkeley Transportation Division, 2019). Parking prices in these programs vary by time of day and location, which, depending on the program, may be by street link or larger urban area. In each pricing unit, the prices are updated periodically until the occupancy rate converges to the desired range, typically between 60 - 80%.

The straightforward way to ensure a uniform occupancy rate is to vary parking prices by small units and indeed, in most of the existing programs, performance prices are varied by street segments or block faces. The prominent example is San Francisco's SFpark pilot program (SFMTA, 2014), where smart sensors were used to measure parking occupancy and establish prices for each individual block. Pricing by block faces or street links is not free of shortcomings. Due to stochastic fluctuations in demand and the small number of spots, minimal parking availability in a unit is still not guaranteed at all times. This undermines the ability of drivers to plan where to park prior to arrival - drivers can decide to park at an affordably priced block face close to their destination, only to find on arrival that it is fully occupied. These drivers then face a complicated decision-making process of searching for parking in a heterogeneously priced parking space whilst under time pressure, which can lead to frustration and illegal behavior.

Prices established over larger areas are easier for drivers to comprehend and abide by. Despite the lack of research on the topic, the cities of Seattle (SDOT, 2019), Calgary (CPA, 2019) and Auckland (AT, 2019) followed this view and established performance parking price by zones that consist of several or more street segments with similar demand profiles. The risk associated with pricing by large units is the high chance for mismatch between local demand and supply, and the emergence of islands of under- and over-occupied parking within the unit.

Performance parking prices require expensive measurements of parking occupancy in every street link (LADOT, 2019; SFMTA, 2014) and periodical price updates until the target occupancy is reached. We present ParkSage, a set of spatially-explicit algorithms for establishing and evaluating parking prices that guarantee a predetermined occupancy rate based on standard GIS layers of residential, commercial and office buildings, parking lots, streets and on-street parking spots. ParkSage continues the tradition of the spatially explicit parking models PARKAGENT and PARKFIT (Levy et al., 2013; Levy and Benenson, 2015) and is free for download at https://www.researchgame.net/profile/Nir_Fulman.

## 2  ParkSage structure

ParkSage consists of three algorithms:

- An algorithm for approximating the parking occupancy pattern given heterogeneous demand and supply patterns (Levy and Benenson, 2015; Fulman and Benenson, 2018[a])
- An algorithm for establishing a pattern of parking prices that guarantees a predefined occupation rate (Fulman and Benenson, 2018[b])
- An algorithm for approximating parking search time based on parking occupancy (Fulman and Benenson, 2018[a])

### 2.1  The algorithm for approximating parking occupancy

Urban parking demand is mostly determined by building capacity and use for residential, commercial, office and other purposes. In addition, parking is necessary near non-building attractions such as parks, open-air markets and historical sites. The algorithm for approximating parking occupancy constructs a *Maximally Dense Parking Pattern (MDPP)* based on urban parking demand at resolution of a single building or attraction, referred to here as a "destination." Destination attributes, such as the floor area of a residential or office building, determine its attractiveness and consequently the number of drivers who arrive to the area and search for parking nearby.

Parking supply in the city is not less heterogeneous than demand, and varies by street links, surface lots and aboveground/underground garages that vary in their capacity, limitations and price. Typically, high-resolution data on demand and supply are available as part of standard municipal spatial databases, and can be converted into spatially explicit demand and supply maps. A building's use, footprint and height can serve for estimating the number of residents and employees and, accounting for car ownership in the area, is sufficient for estimating the demand for parking by hour of the day and day of the week. GIS layers of parking lots typically contain information on their capacity, which can be otherwise estimated based on their area. In case information about on-street parking is unavailable, curb parking spots can be constructed programmatically, 5 meters apart from each other on both sides of a two-way street link, and on the right side of a one-way link, with some shift from the junction. Parking regulations and restrictions, if available, can also be accounted for.

The MDPP algorithm generates a parking occupancy pattern based on spatially explicit high-resolution demand and supply information. As reflected in its name, the MDPP supplies a parking pattern that is "maximally dense," whereas real drivers possibly exploit parking space less effectively. Informally, the MDPP algorithm manages parking in a city of autonomous vehicles that are governed by one control station possessing complete information on the destinations of all arriving vehicles, and all occupied and vacant parking places. When a vehicle arrives to the area, the system reserves the vacant parking spot that is most desirable to the driver in terms of price and walking time from parking place to destination. Vehicles are fully controlled by the system and, thus, a reserved spot can be occupied by the assigned vehicle only.

Formally, let $A_{c,p}(d)$ be the attractiveness of a parking spot p at a distance d from the driver's c destination, and $F_p$ be the parking price of p. We assume that $A_{c,p}(d)$ satisfies two conditions:

- $A_{c,p}(d)$ decreases with the increase in distance d between p and the destination as $1/d^\alpha$,

$$A_{c,p}(d) \sim 1/d^\alpha, \text{ where } 0 < \alpha < 1 \qquad (1)$$

- Driver c is sensitive to the parking price $F_p$ only when $F_p$ is above a threshold "negligible" level $f_{negligible} > 0$, and $A_{c,p}(d)$ depends on $F_p$ only when $F_p > f_{negligible}$. Formally, we express this as

$$A_{c,p}(d) = \min(1, f_{negligible}/F_p)/d^\alpha, \text{ where } 0 < \alpha < 1 \qquad (2)$$

Based on parking studies in the field (Levy and Benenson, 2015), we assume that an acceptable walking distance d between parking spot and destination is limited to 500m aerial distance, $d_{max} < 500m$.

We also assume that drivers whose parking options all have low attractiveness, may give up on parking. Let $A_{c,best}$ be the attractiveness of the best available parking spot for the driver c. The probability $g(A_{c,best})$ to give up on parking is calculated as:

$$g(A_{c,best}) = \begin{cases} 0 & if\ A_{c,best} > A_{threshold} \\ 1 - \exp(\gamma(1 - A_{threshold}/A_{c,best})) & if\ A_{c,best} \leq A_{threshold} \end{cases} \qquad (3)$$

where γ is a parameter. In all examples below γ = 0.1.

Drivers who fail to find parking within distance $d_{max}$ from destination or forgo parking due to low attractiveness of the available spots, are assumed to park far away and use additional means of transportation to reach their destinations.

Formally, the MDDP algorithm can be represented by the following flow charts (Fulman and Benenson, 2018[a]):

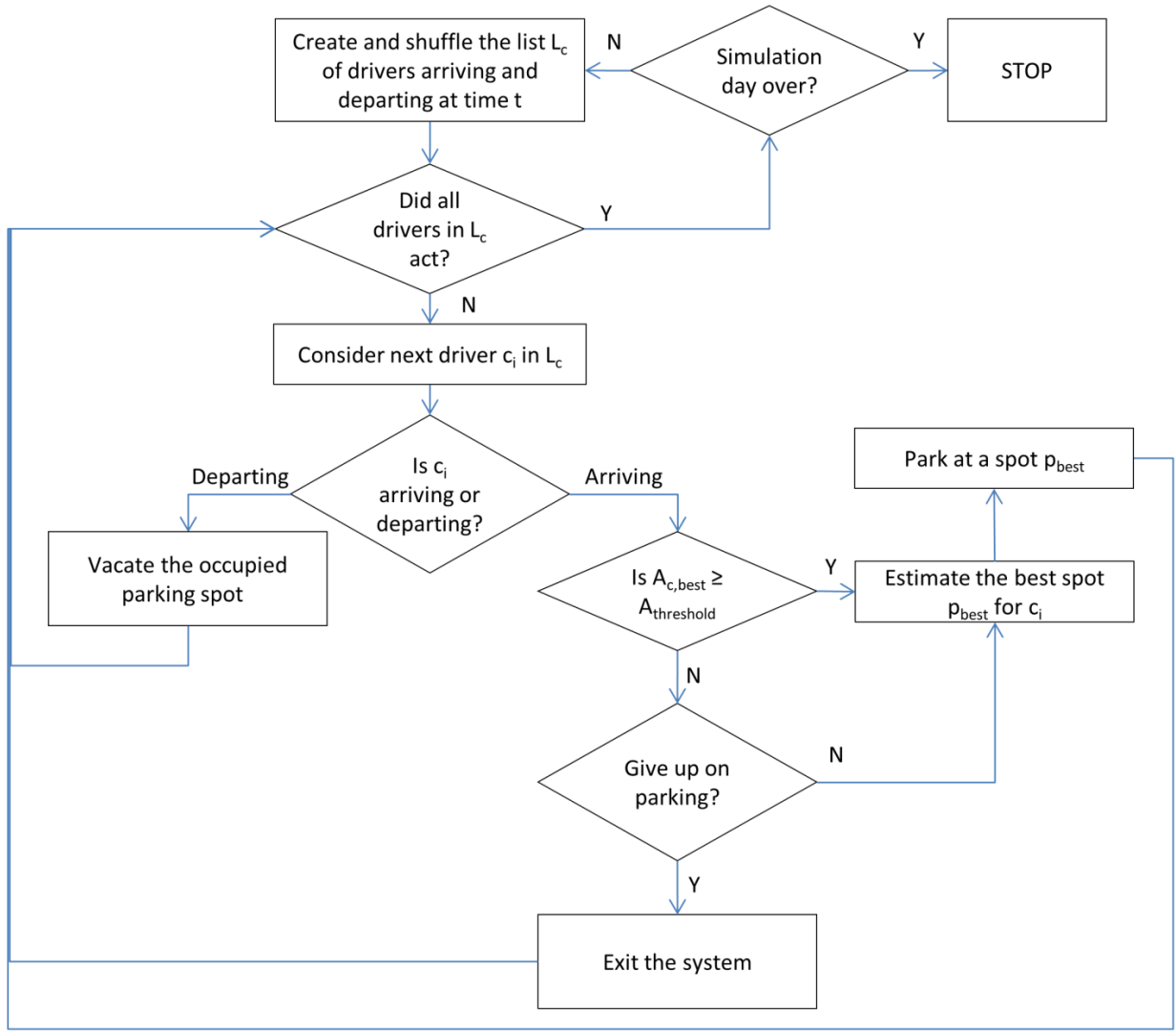

Figure 1. Flow chart of the Maximally Dense Parking Pattern (MDPP) algorithm.

## 2.2  An algorithm for establishing demand-responsive parking prices

The Nearest Pocket for Prices Algorithm (NPPA) applies the MDPP in a recursive way for approximating parking prices. Let the target occupation level in the area be $O_{threshold}$. The idea is to increase the price $F_u$ of each parking unit u (e.g. block face or neighborhood), where the average occupancy exceeds the threshold $O_{threshold}$, from $F_u$ to $F_u(1 + \phi)$, and to redistribute the drivers who arrive to park in the area, by applying the MDPP with the new price pattern. Increasing the parking price in a unit reduces the attractiveness of parking places in it, and drivers who are sensitive to the new price respond by preferring to park elsewhere, typically farther away from their destination, where parking is cheaper. The algorithm quits when the price of parking at each unit u reaches a level that guarantees an average occupancy rate $O_u \leq O_{threshold}$. The NPPA's flow is as follows (Fulman and Benenson, 2018[b]):

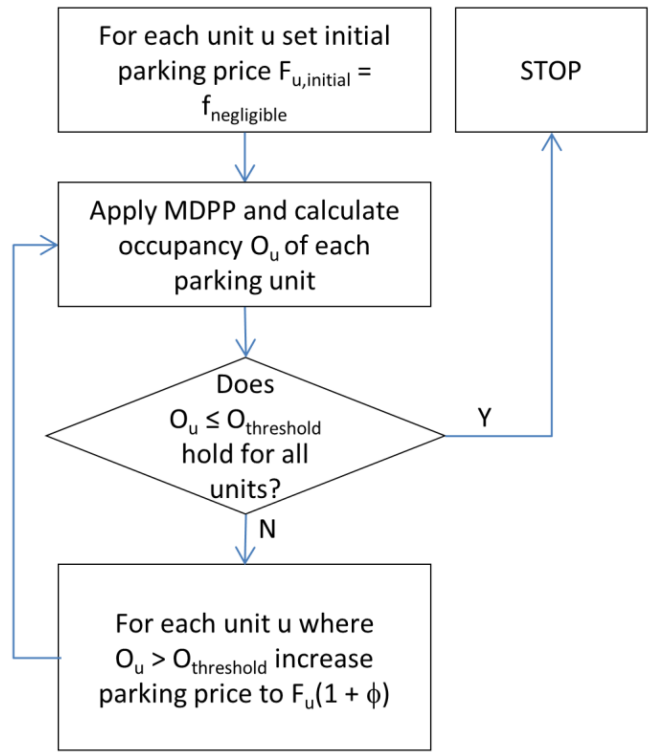

Figure 2. Flow chart of the Nearest Pocket for Prices Algorithm (NPPA) algorithm.

Below we apply the NPPA algorithm with the $\phi = 0.05$. The typical number of iterations necessary for convergence to a steady price pattern is 50.

## 2.3 An algorithm for estimating cruising time based on the occupation pattern

Parking search time increases with the increase in the occupation rate in the driver's search area. However, the average occupancy rate over a spatial unit does not directly translate to parking availability and cruising time, due to the spatial heterogeneity of demand and supply within the unit. Larger spatial units contain many street links and buildings and, given the average occupation rate $O_u$ over the entire unit is below the preferred $O_{threshold}$, the number of links within it that are occupied at a level that exceeds $O_{threshold}$, increases with the size of the unit. Since parking is impossible on fully occupied links, their occurrence within the destination's search neighborhood determines cruising time.

The algorithm for predicting parking search time distribution for the driver whose destination is n is based on the distribution of link occupancy within the destination's search neighborhood. It builds on two components: The MDPP-provided occupation rate $p(l)$ for every link $l \in N(n)$ (Fulman and Benenson, 2018[a]) and experimentally discovered probability $w(l, n)$ of including each link l ∈ N(n) in a cruising path, depending on the distance between l and n (Fulman and Benenson, 2019). Given the patterns of w(l, n) and p(l) within N(n), and assuming that a driver traverses one link per time unit, we are able to estimate probability $q_{nopark}$ to traverse a fully occupied link per time unit (Fulman and Benenson, 2018[a]):

$$q_{nopark} = \sum_{l \in N(n), l\ is\ fully\ occupied} w(l,n)p(l) / \sum_{l \in N(n)} w(l,n)p(l) \quad (4)$$

The probability Q(τ, n) to cruise longer than τ time units, for the driver whose destination is n, is:

$$Q(\tau, n) = (1 - q_{nopark})q_{nopark}^{\tau} \quad (5)$$

The flow chart of the algorithm for estimating cruising time is as follows (Fulman and Benenson, 2018[a]):

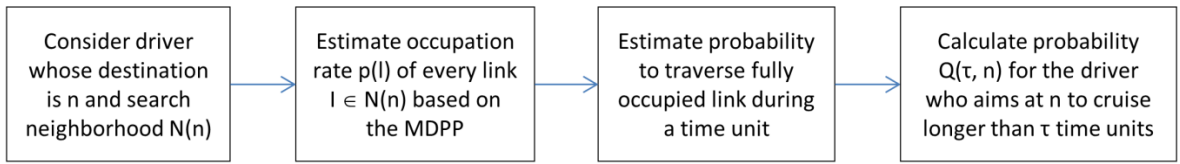

Figure 3. Flow chart of the algorithm for estimating cruising time distribution.

The set of ParkSage algorithms is sufficient for establishing and evaluating the efficiency of various pricing schemes in a city. In what follows, we apply ParkSage for establishing overnight parking prices in the Israeli city of Bat Yam, comparing three methods of partitioning the city into priced units.

## 3    Establishing parking prices in the city of Bat Yam

We apply ParkSage for establishing overnight parking prices in Bat Yam, using data from 2010 (Martens et al., 2010). The total Bat Yam population in 2010 was ca. 130,000, car ownership 35,000, and the number of residential buildings 3,300 with 51,000 apartments. Residential buildings in Bat Yam provide their tenants a total of 17,500 dedicated parking places that we exclude from the demand and supply data.

### 3.1    Bat Yam demand and supply

Parking supply in Bat Yam data is given in two GIS layers - a layer of streets and a layer of off-street parking facilities. Based on the layer of streets, 27,000 spots for curb parking were constructed automatically, 5 meters apart on both sides of two-way street links, and on the right side of one-way links, with a necessary gap from the junction. In addition, 1,500 spots are available for the city's residents in its parking lots, where Bat Yam residents can park in the evening free of charge. The average overnight demand/supply ratio is thus very low (35,000 – 17,500) / (27,000 + 1,500) ≈ 0.61 car/parking spot. However, the distributions of demand and supply in Bat Yam are both highly heterogeneous, and the overnight demand in the center of Bat Yam significantly exceeds the supply there (Figure 4).

## 3.2 Arrivals and departures

In the experiments below, we consider evening (16:00 to 23:00) on-street parking by residents and their guests. We assume that all parking spots are vacant at 16:00 and residents arrive to the area between 16:00 and 18:00 and park until the end of the evening. Guests arrive and depart throughout the whole evening, and their parking time is uniformly distributed on the [$\tau_{min}$, $\tau_{max}$] interval, where $\tau_{min}$ = 1 hour and $\tau_{max}$ = 2 hours.

Drivers that aim at a destination $n_i$ are generated by a Poisson process with a per-hour average $\lambda_i$ based on the $n_i$'s hourly demand $d_i$ by residents and guests. We assume that for each $n_i$, residents comprise a constant fraction e < 1 of the total demand $d_i$. During that time period, we set the hourly number of arrivals to each destination $n_i$ as $\varepsilon_i$ = e*$d_i$/2. We adjust the hourly arrival rate of guests, to guarantee that the average number of guests arriving to $n_i$ be equal to (1 − e)*$d_i$. Accounting for the guests' average parking time ($\tau_{min}$ + $\tau_{max}$)/2, their per hour arrival rate to a destination $n_i$ is $\lambda_i$ = 2*(1 - e)*$d_i$/($\tau_{min}$ + $\tau_{max}$), and we assume that it remains constant throughout the evening.

At each 30 seconds model tick, the list of arriving and departing drivers is created, randomly re-ordered, and each driver acts in its turn, facing the parking pattern created by the actions of its predecessors. Average occupation rate in the city converges to equilibrium towards 20:00 and then fluctuates, very slightly, over time.

## 3.3 Cruising for parking in Bat Yam with the existing free of charge parking

Let us estimate evening cruising time in Bat Yam, assuming that the ratio between residents and guests seeking parking is 85:15, and that parking on-street and in public lots is free to residents of Bat Yam whereas outsiders are required to pay a yearly fee of 300 ILS for a parking permit. We consider the permit price as the negligible threshold of locals (see formula (2)) and establish $f_{negligible}$ per night as $f_{negligible}$ = 1 ILS, roughly equivalent to $0.3. In what follows, we refer to this price as the base price.

The patterns of supply and residential demand for overnight parking at a resolution of buildings, street links and lots, and the demand to supply ratio aggregated over Transport Analysis Zones (TAZ), are presented in Figure 4.

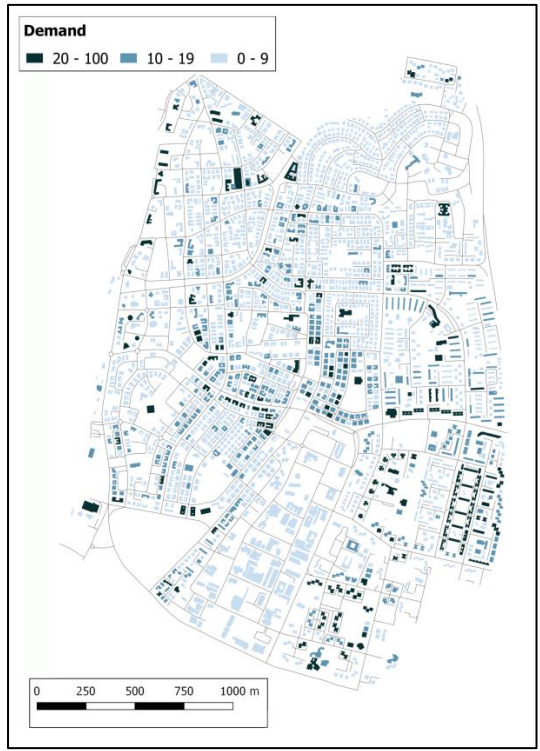
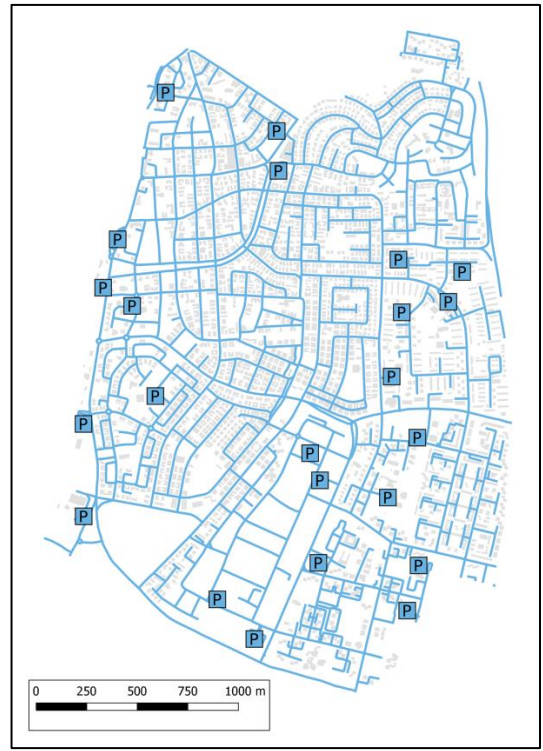
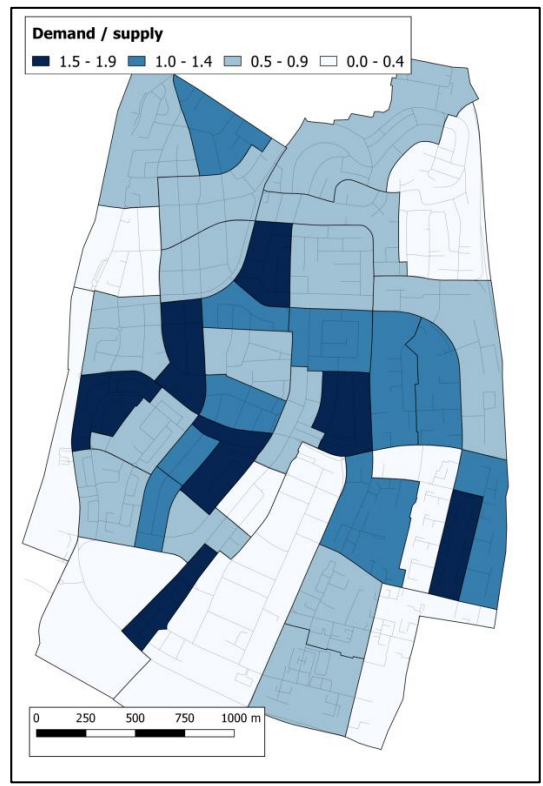

Figure 4. Bat Yam: (a) Parking demand by buildings; (b) Parking supply by street links and of-street lots; (c) Demand-to-Supply ratio by TAZ.

Drivers aim to park as close as possible to their destinations. Where night parking is free, Bat Yam drivers clog up links within the residential blocks where demand exceeds supply and spill over to adjacent areas, worsening parking conditions there. As a result, in Bat Yam's center, parking in the evening becomes unavailable over large continuous areas. The resulting cruising time is long: The estimated average cruising time for most destinations in and around the city center is, on average, 2.5 minutes; and cruising for longer than 5 minutes becomes very common (Figure 5).

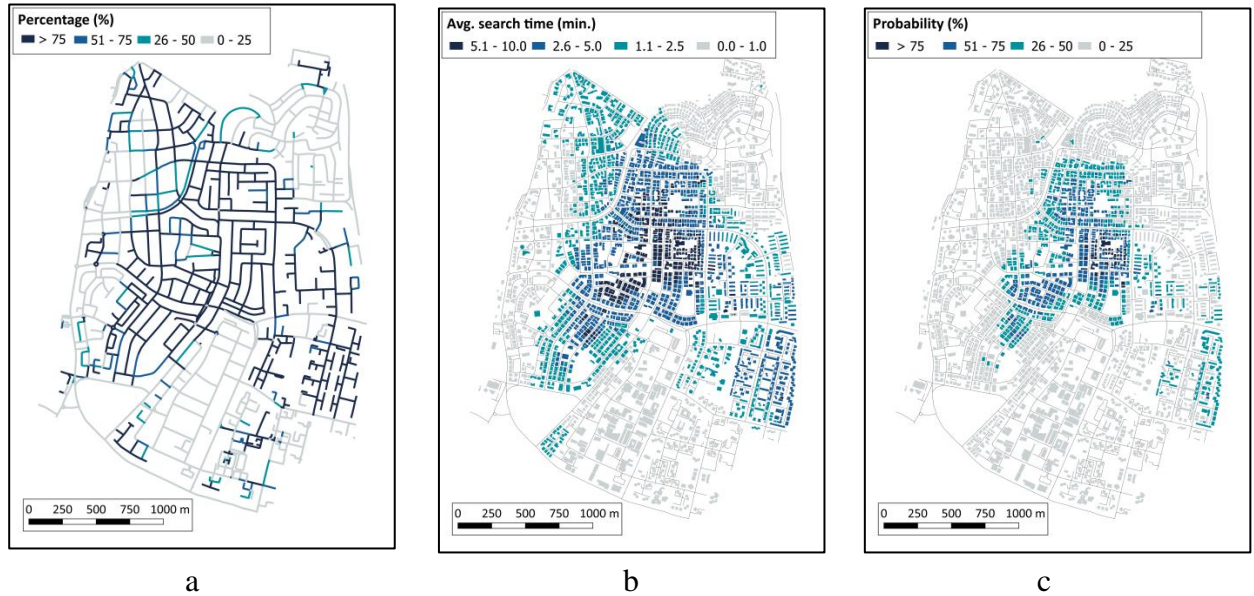

Figure 5. The percentage of time that street segments are fully occupied (a); the average parking search times (b); and the probability to cruise for over 5 minutes (c) in Bat Yam.

### 3.4 Establishing parking prices that guarantee close-to-zero cruising time

We establish parking prices for the traditional threshold occupation $O_{threshold}$ = 85% (1 of 7 spots is vacant) in each street segment, commonly accepted as guaranteeing close-to-zero cruising (Shoup, 2006). In these model experiments we assume that the parking fee is charged once for the entire evening and parking durations are not influenced by the prices. We also assume that residents and guests are equally sensitive to prices, the distance-payment tradeoff coefficient in formula (1) is $\alpha$ = 0.5, the coefficient $\gamma$ in formula (2) is $\gamma$ = 0.1 and the minimal attractiveness threshold $A_{threshold}$ = 0.1.

Despite a global demand to supply ratio far below 1, demand in the center of Bat Yam is essentially higher than supply (see Figure 4 above). As a result, to maintain an average occupancy rate of 85%, prices in the busiest blocks should be 25-30 ILS, whereas at the periphery of the city the prices can remain at the base level (Figure 6). Similar to the outcomes of the SFpark experiment (SFMTA, 2014), one can frequently observe a twofold difference between parking prices for two adjacent street links.

The established pattern of parking prices ensures that almost none of the street links is fully occupied for longer than an hour and clusters of two or more connected links that are fully occupied for longer than half an hour almost never emerge. With parking available on every link almost all of the time, cruising time is reduced nearly to zero. The average parking search time over all destinations is shorter than 60 seconds and less than 1% of drivers are expected to cruise for longer than 150 seconds.

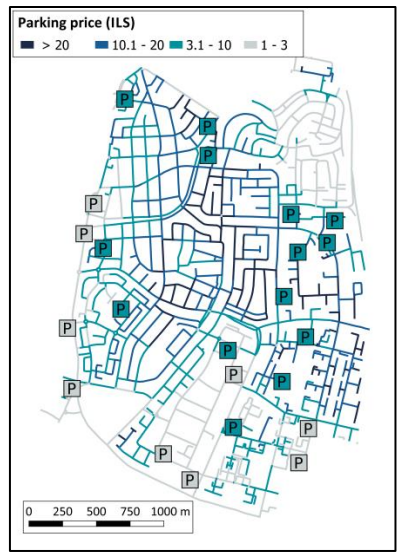

Figure 6. Equilibrium parking prices, in ILS, for 85% occupation threshold established by street links and parking lots as pricing units.

## 3.5 Pricing parking by large spatial units

We establish parking prices for coarse partitions that are convenient for drivers, starting with Bat Yam's 6 largest administrative units, referred to as quarters (Figure 7). Each quarter contains between 2700 and 5800 parking places, and their size varies between 0.7-1.4 square km. The application of ParkSage with $O_{threshold} = 85\%$ results in upper-than-base prices in one quarter only - the core of the Bat Yam city, which is about 1 km$^2$ and contains over 3000 parking places. The parking price in this core quarter is 8.5 ILS (~$2.5).

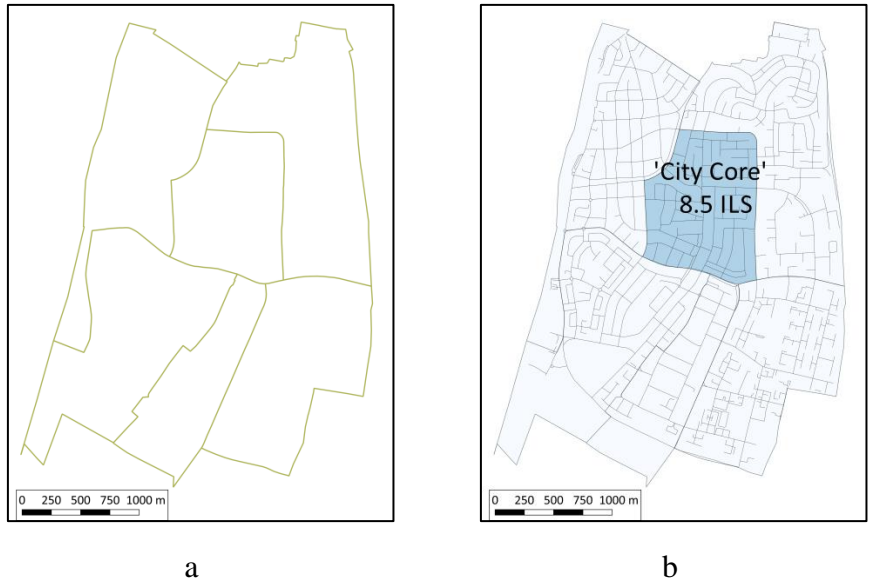

Figure 7. Six quarters of Bat Yam (a); and equilibrium parking prices, in ILS, for 85% occupation threshold estimated by city quarters as pricing units (b).

Priced parking in the central quarter drastically reduces cruising for almost all destinations there and for some external destinations close to its border. The difference between cruising time for the case of no pricing versus priced parking in the central quarter is close to 3 minutes on average and 75% of the drivers whose destinations are in this quarter cruise for less than 90 seconds.

Yet pricing the central quarter does not guarantee parking availability everywhere in it. Priced parking there pushes some drivers to park in adjacent quarters, while the occupation rate of some links within the core remains far above 85% (Figure 8). Similarly, in the quarters where prices aren't established, clusters of fully occupied links emerge around the locations where demand exceeds supply and, in addition, areas bordering on the central quarter (Figure 9). The problem is thus partially displaced instead of being resolved. In Bat Yam, large and heterogeneous spatial units of 1 km$^2$ are inappropriate for pricing because the average occupancy rate does not adequately represent parking availability in the city.

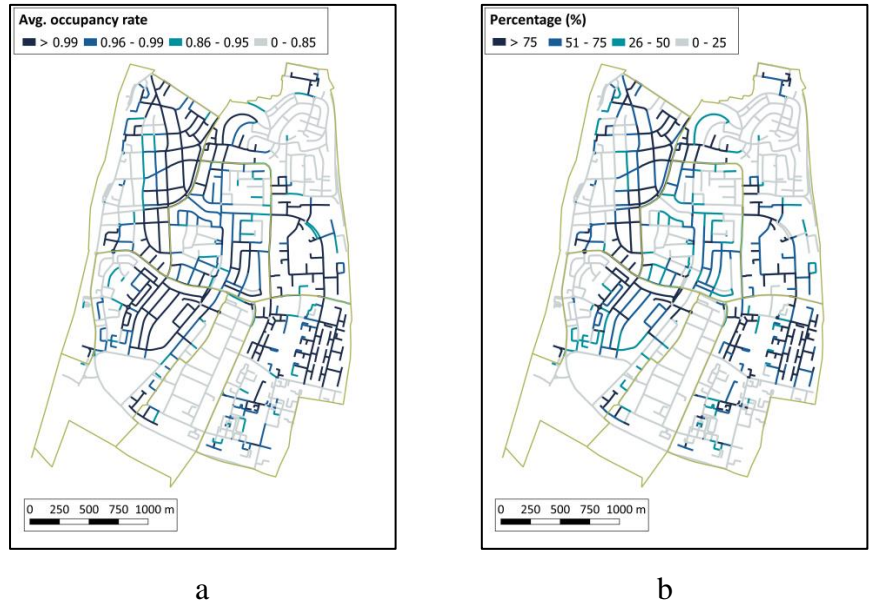

Figure 8. Bat Yam parking pricing by quarters: Average occupancy rate by street segments (a); percentage of time that street segments are fully occupied (b).

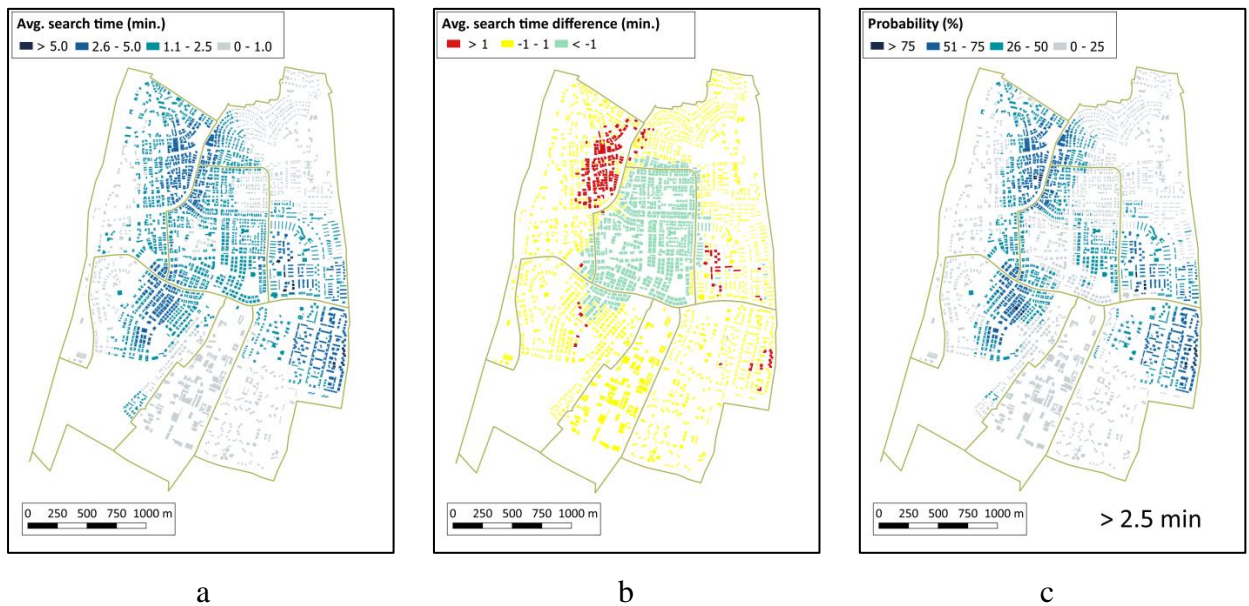

Figure 9. Parking search times in Bat Yam for pricing by city quarters: Average parking search times (a); average difference in search times when compared to non-priced parking (b); and probability to cruise longer than 2.5 minutes (c).

Let us investigate the consequences of pricing by the intermediately coarse partition of Bat Yam into 42 Transport Analysis Zones (TAZ). The average TAZ is 0.16 square km with an average parking capacity of 600 spots, making a TAZ substantially larger than a street link and smaller than a quarter. Figure 10 presents TAZ-based parking prices that guarantee 85% occupation rate.

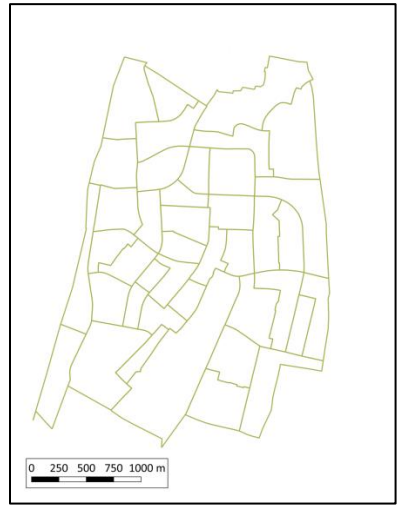 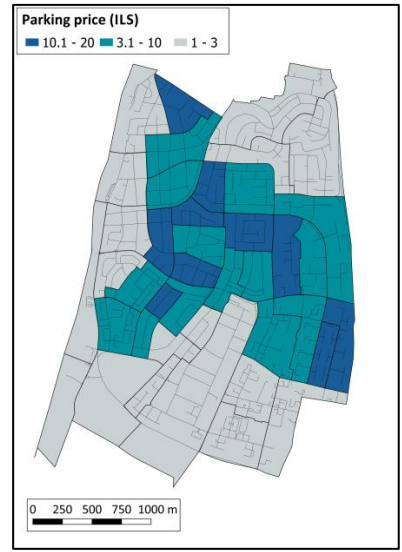

a  b

Figure 10. The 42 transportation analysis zones (TAZ) in Bat Yam (a); equilibrium parking prices, in ILS, for 85% occupation threshold based on TAZ as pricing units (b).

The prices are above the base level in 24 TAZ, all located in and around the center of the city, and prices range up to 17.5 ILS ($5).

Despite being essentially smaller than quarters, pricing by TAZ does not guarantee 85% occupation for every single link. In many links parking in the evening is still unavailable, and their average occupancy rate over the evening remains close to 100%. However, with TAZ-based pricing, overly-occupied links rarely form large spatial clusters and parking is always available one or two links away. As a result, the average search time does not exceed 2.5 minutes for any destination and the average cruising time for the 500 destinations around which visitors cruise the longest, is only 75 seconds (Figure 11).

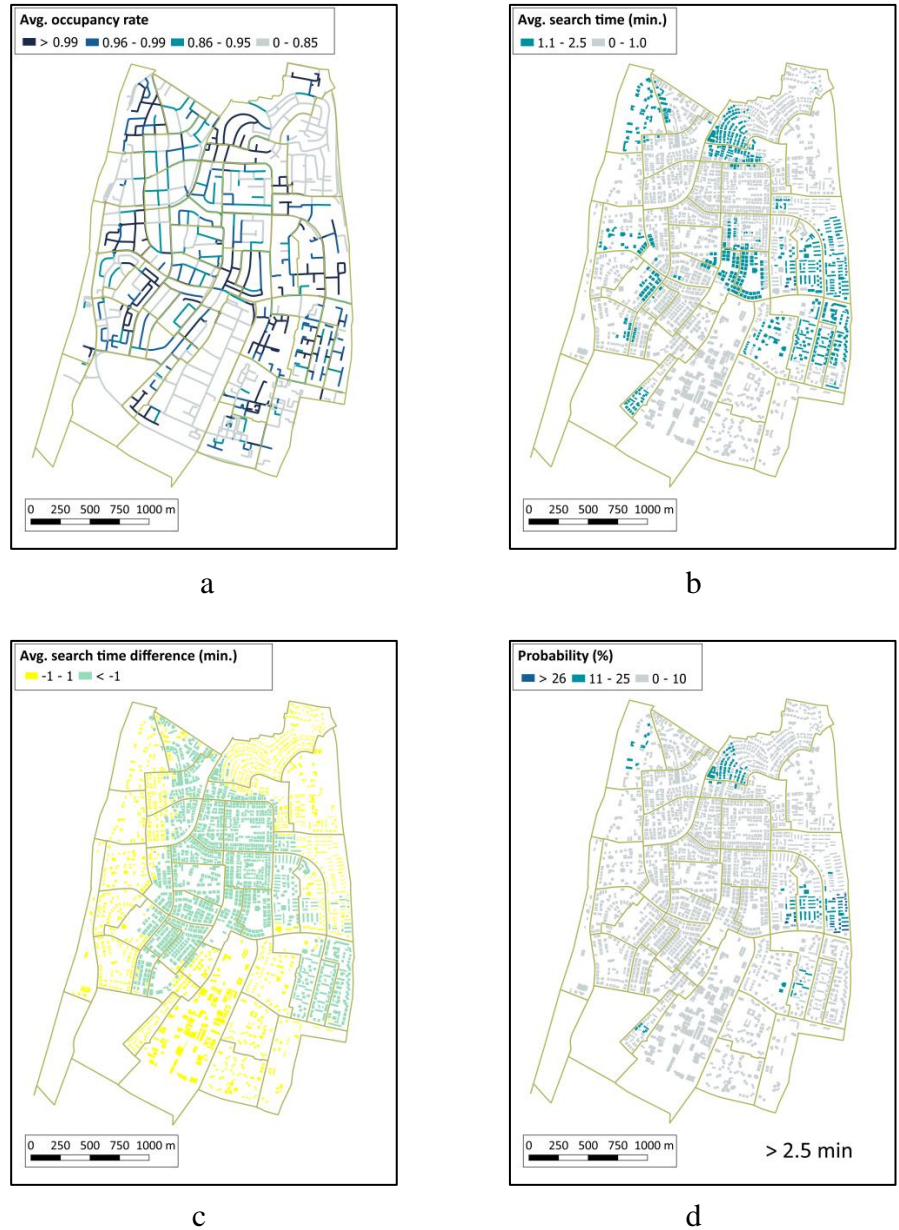

Figure 11: Characteristics of Bat Yam TAZ-based pricing: Average occupancy rate by links (a); average parking search time (b); average difference in search times compared to non-priced parking (c); probability to cruise longer than 2.5 minutes (d).

## 4 The choice of optimal partition for demand-responsive pricing

In most cities around the globe, parking prices are controlled by law. Prices were generally established a long time ago and remain steady to avoid protest from car owners and politically costly public debate. As a result, a substantial fraction of drivers is exempt from paying, the others pay less than the situation requires, and all drivers suffer from long and exhausting searches for parking. Performance pricing of parking aims to balance parking demand and supply. The shift to market-defined prices is of dire importance, and establishing the minimal prices that would improve parking conditions demands solid guidelines since one design flaw may sabotage an otherwise smart policy. The use of ParkSage, a spatially explicit toolset for establishing demand-responsive parking prices, can assist in ensuring the reliability of performance pricing.

Practical experiments for establishing performance-defined pricing schemes have had only limited success (SFMTA, 2014; Pierce and Shoup, 2013; Chatman and Manville, 2014; Millard-Ball et al., 2014). It is hypothesized that the constraints placed upon these parking programs, including lax enforcement of illegal parking, free parking permits, abuse of disabled parking placards and price ceilings are to blame for this. Our work shows that unconstrained performance pricing should be able to reduce parking search substantially, if not end it completely.

The basic rule of thumb in performance pricing is to keep 15% of parking places vacant in each link or block, guaranteeing one of seven spots open on each block face. However, due to stochastic variations of demand, parking can never be guaranteed on every block all of the time. Searching for the ideal place when parking prices vary by the block is challenging, and may lead to frustration, prolonged cruising, and encourage illegal behavior. Pricing by units larger than street links is simpler to conduct and easier to convey to drivers. We show that pricing by medium sized units, such as Transport Analysis Zones, is sufficient for preserving an occupation level that prevents cruising for parking.

Using ParkSage for examining the effects of pricing by unit size in Bat Yam, we have demonstrated that pricing by the very large city quarters does not prevent substantial cruising by visitors. It also brought to light a potential caveat of confining performance pricing to neighborhoods notorious for long cruising, as done in the pilot areas in San Francisco's SFpark program (SFMTA, 2014) and Washington D.C.'s Chinatown and Penn Quarter (DDOT, 2019). In these pilot areas, drivers whose destinations are within a pricing unit yet close to its borders often search for cheaper parking beyond it and these spillovers worsen parking conditions in adjacent areas, introducing cruising for parking where it did not previously exist.

ParkSage can be applied not only for establishing optimal pricing partitions, but in several additional ways, one of which is assessment of parking constraints and permissions. To date, all performance pricing programs that we are aware of, including the prominent SFpark project, are constrained in one or more ways. Illegal parking, parking permits, disabled parking placards and other limitations can be modeled in ParkSage by taking into consideration drivers who are completely or partially insensitive to parking prices. Using ParkSage in this way, one can reveal where and when it is plausible to expect good results despite the constraints and permissions, and where they may render pricing ineffective.

Time-varying pricing, including time-varying pricing units, is another aspect of performance pricing that can be explored using ParkSage. Inclusion of time into the pricing policy demands an additional compromise: If the price remains uniform over a period of time that includes inherent variations in demand, the average occupancy rate over that period does not represent parking availability. On the other hand, very short time bands would be equally as frustrating for drivers as prices that vary by street links. We did not investigate time-varying prices in this paper, but it should be noted that given sufficient information about the daily variation of demand, ParkSage can be used for choosing the appropriate time bands, especially for establishing daily intervals of charged and free parking.

## Acknowledgement

This research was supported by the ISRAEL SCIENCE FOUNDATION (grant No. 1160/18) "Tessellation of urban parking prices".